\documentclass[12pt,notitlepage,tightenlines,eqsecnum,floats,showpacs,nofootinbib,amsmath,amssymb,aps,prd]{revtex4-2}

\usepackage{amsmath,amssymb,amsfonts,latexsym,cancel}
\usepackage{mathtools}
\usepackage{graphicx}
\usepackage{color}
\usepackage{hyperref}
\usepackage[margin=0.8in]{geometry}
\usepackage{tikz}
\usepackage{natbib}

\newcommand{\be}{\begin{equation}}
\newcommand{\ee}{\end{equation}}

\newcommand{\bma}{\begin{bmatrix}}
\newcommand{\ema}{\end{bmatrix}}

\newcommand{\pd}{\partial}

\def\ea{E^a}
\def\eb{E^b}
\def\f{\frac}
\def\nn{\nonumber}
\def\lp{\ell_{\rm Pl}}

\begin{document}

\title{Effective loop quantum gravity framework for \\ vacuum spherically symmetric space-times}

\author{Jarod George Kelly} \email{jarod.kelly@unb.ca}
\affiliation{Department of Mathematics and Statistics, University of New Brunswick, Fredericton, NB, Canada E3B 5A3}

\author{Robert Santacruz} \email{robert.santacruz@unb.ca}
\affiliation{Department of Mathematics and Statistics, University of New Brunswick, Fredericton, NB, Canada E3B 5A3}

\author{Edward Wilson-Ewing} \email{edward.wilson-ewing@unb.ca}
\affiliation{Department of Mathematics and Statistics, University of New Brunswick, Fredericton, NB, Canada E3B 5A3}

\begin{abstract}

We develop an effective framework for the $\bar\mu$ scheme of holonomy corrections motivated by loop quantum gravity for vacuum spherically symmetric space-times.  This is done by imposing the areal gauge in the classical theory, and then expressing the remaining components of the Ashtekar-Barbero connection in the Hamiltonian constraint in terms of holonomies of physical length $\lp$.  The stationary solutions to the effective Hamiltonian constraint can be found exactly, and we give the explicit form of the effective metric in Painlev\'e-Gullstrand coordinates.  This solution has the correct classical limit, the quantum gravity corrections decay rapidly at large distances, and curvature scalars are bounded by the Planck scale, independently of the black hole mass $M$.  In addition, the solution is valid for radii $x \ge x_{\rm min} \sim (\lp^2 M)^{1/3}$ indicating the need for a matter field, with an energy density bounded by the Planck scale, to provide a source for the curvature in the space-time.  Finally, for $M \gg m_{\rm Pl}$, the space-time has an outer and also an inner horizon, within which the expansion for outgoing radial null geodesics becomes positive again.  On the other hand, for sufficiently small $M \sim m_{\rm Pl}$, there are no horizons at all in the effective metric.

\end{abstract}
\maketitle

\section{Introduction}

It is widely expected that the singularity at the center of a black hole, predicted by classical general relativity, signals the breakdown of the classical theory and indicates the need to include quantum gravity effects.  In addition, the black hole information loss problem further underlines the importance of understanding the role of quantum gravity effects in black hole space-times, with modifications to the causal structure of the quantum-corrected space-time being especially relevant.

One approach to explore how quantum gravity could affect black hole space-times is by modifying the metric of the Schwarzschild space-time, often by hand, to provide concrete examples of non-singular space-times \cite{Bardeen:1968, Frolov:1981mz, Roman:1983zza, Dymnikova:1992ux, Borde:1996df, AyonBeato:1998ub, Dymnikova:2003vt, Hayward:2005gi, DeLorenzo:2014pta, Frolov:2016pav, Simpson:2019mud}, that can then be classed according to their geometry \cite{Carballo-Rubio:2018jzw, Carballo-Rubio:2019fnb}.  However, in principle the best would be to start from a specific theory of quantum gravity, to determine the states that correspond to spherically symmetric space-times, and to extract physical predictions from these states.  While this so far remains an outstanding challenge for all candidate theories of quantum gravity, it is nonetheless possible to include certain effects---predicted by particular theories---and study their impact on black hole space-times.  There has been considerable work in this direction in a number of quantum gravity theories (see, e.g., \cite{Mathur:2005zp, Nicolini:2005vd, Falls:2010he}), and in particular in loop quantum gravity (LQG), a background-independent and non-perturbative theory of quantum gravity \cite{Ashtekar:2004eh, Rovelli:2004tv, Thiemann:2007pyv}.

The work in LQG has built on earlier research that studied LQG effects in cosmological space-times following the loop quantum cosmology (LQC) procedure: first, the symmetries of the space-time of interest are imposed at the classical level, and second, the symmetry-reduced classical theory is quantized using LQG methods---notably, the fundamental operators are holonomies of the connection and areas.  For a review on LQC, see, e.g., \cite{Ashtekar:2011ni}; this same procedure has since been applied to black hole space-times as well.

There has been a considerable focus on the Schwarzschild interior, using the isometry between the interior and the Kantowski-Sachs space-time to more easily import techniques from LQC \cite{Modesto:2004xx, Ashtekar:2005qt, Bohmer:2007wi}. Despite this effort, as shall be discussed in Sec.~\ref{sec:HC} in more detail, it has turned out to be difficult to handle holonomy corrections properly in this framework.  To ensure that the edges along which the holonomies are evaluated have a physical length $\sim \lp$, it is necessary to relate the physical length to a coordinate length by the metric; this is called the `$\bar\mu$ scheme'.  However, it is not clear how to properly take into account the $\bar\mu$ scheme near the horizon when using a set of coordinates where a spatial coordinate becomes null at the horizon and the physical length along that coordinate tends to 0, as is the case for the Schwarzschild interior in Kantowski-Sachs coordinates \cite{Bohmer:2007wi}.  There exist various proposals in the literature to address this difficulty \cite{Campiglia:2007pb, Chiou:2008nm, Brannlund:2008iw, Joe:2014tca, Corichi:2015xia, Cortez:2017alh, Olmedo:2017lvt, BenAchour:2018khr, Ashtekar:2018cay, Bodendorfer:2019cyv, Alesci:2019pbs, Assanioussi:2019twp}, but here we suggest instead that the $\bar\mu$ scheme simply cannot be implemented in terms of a particular set of (spatial) coordinates if one of these coordinates becomes null, for example at a horizon.

There has also been significant work studying the dynamics of the full space-time---interior and exterior together---with LQG-inspired corrections, whether holonomy effects \cite{Bojowald:2005cb, Gambini:2008dy, Gambini:2013hna, BenAchour:2016brs, Bojowald:2018xxu, Reyes:2009, Gambini:2013ooa} or inverse triad effects \cite{Husain:2004yz, Ziprick:2009nd, Bojowald:2009ih, Kreienbuehl:2010vc, Bojowald:2011js}, and there has also been some more recent work in this setting where it has been shown how to implement the $\bar\mu$ scheme \cite{Chiou:2012pg, Gambini:2020nsf}, which raises the hope that by considering the whole space-time at once it may be possible to avoid the difficulties that arise when considering the Schwarzschild interior only.

In this paper, we will further study the $\bar\mu$ scheme for holonomy corrections in vacuum spherically symmetric space-times and also extend earlier results in a manner so that the extension to include matter fields will be quite direct; in particular, it is straightforward to include pressureless dust \cite{dust}.  For previous work on including matter in spherically symmetric space-times (although not in the $\bar\mu$ scheme), see \cite{Reyes:2009, Gambini:2009ie, Gambini:2014qta, Bojowald:2015zha, Campiglia:2016fzp}.

By including matter, it is possible to study how quantum gravity effects may arise during black hole collapse, and how they could modify the resulting space-time.  LQG effects on black hole collapse have previously been explored in a number of settings \cite{Husain:2006cx, Husain:2008tc, Ashtekar:2008jd, Hossenfelder:2009fc, Ashtekar:2010qz, Tavakoli:2013rna, Christodoulou:2016vny, Christodoulou:2018ryl, Benitez:2020szx, BenAchour:2020gon}, and extending these studies to include the $\bar\mu$ treatment of the holonomies in a way that provides a general framework that determines the dynamics for both the interior and exterior regions will set the stage for more detailed investigations into the role of quantum gravity effects on gravitational collapse.

In particular, one possibility that has been suggested is that when the energy density of the matter composing the collapsing star reaches the Planck scale, quantum gravity effects could generate a non-singular transition to a slowly expanding white hole solution \cite{Rovelli:2014cta, Haggard:2014rza, Barcelo:2014cla, Bianchi:2018mml}, with potential observational implications \cite{Barrau:2014hda, Barrau:2014yka, Barrau:2015uca}.  It turns out that this general picture is explicitly realized in this framework when a pressureless dust field is coupled to gravity; for details see \cite{dust}.

We begin the paper with a general discussion on holonomy corrections and some specific comments on the difficulties that arise in black hole space-times in Sec.~\ref{sec:HC}, then present the classical Hamiltonian framework and impose the areal gauge in Sec.~\ref{sec:CD}, and finally construct the effective theory with LQG holonomy corrections and study its solutions in Sec.~\ref{sec:ED}.  Although the effective theory is obtained by following a different path, the results are in perfect agreement with \cite{Gambini:2020nsf}, showing the robustness of the results.  We end with a discussion in Sec.~\ref{sec:sum}

\newpage

Our conventions are the following: space-time indices are denoted by $\mu, \nu, \rho, \sigma, \ldots$; spatial indices are denoted by $a, b, c, \ldots$; and internal indices are denoted by $i, j, k, \ldots$  We use units where $c=1$, but leave $G$ and $\hbar$ explicit to clarify the interplay of gravitational and quantum effects.

\section{Holonomy Corrections}
\label{sec:HC}

To develop an effective framework for vacuum spherically symmetric black holes following the standard LQC procedure, it is necessary to incorporate holonomy corrections in an appropriate fashion.  In this section, we will briefly review the main steps, offer an explanation on why it has been found to be difficult to properly implement the $\bar\mu$ scheme for holonomy corrections in the Schwarzschild interior, and explain the procedure we will follow in this paper to avoid these difficulties.

\subsection{Holonomies}

The holonomy of the Ashtekar-Barbero connection $A_a = A_a^i \tau_i$ along a path $\ell$ is given by
\be \label{def-hol}
h_\ell(A) = \mathcal{P} \exp \left( \int_\ell A_a \right),
\ee
where $\mathcal{P}$ denotes path-ordering, while the $\tau_i$ are a basis of the $\mathfrak{su}(2)$ Lie algebra.  There are two points here that are important to understand for what follows.

First, the connection is usually expressed in terms of some coordinates, in which case the length of the edge as calculated in terms of these coordinates will necessarily be a coordinate length $L_c$.  So to calculate the holonomy along a path that has a specific physical length $L_p$, it will be necessary to use the space-time metric $g_{\mu\nu}$ to relate the coordinate and physical lengths to calculate the required coordinate length $L_c$.

Second, the path-ordered exponential of an integral is defined by a series of nested integrals which, in general, are typically difficult to evaluate.  In the simple case when the connection is independent of a particular coordinate, then the path-ordering trivializes for holonomies in that direction and \eqref{def-hol} can be evaluated much more directly.  This is relevant for spherically symmetric space-times: holonomies in the radial direction will be difficult to evaluate, while holonomies along paths where only the angular coordinates vary will be much easier to calculate.

\subsection{Black Holes and the Near-Horizon Region}

In LQC, the holonomies are taken along paths of physical length $\sqrt\Delta$, where the area gap $\Delta$ is the minimum non-zero area eigenvalue in LQG \cite{Ashtekar:2006wn}.  As explained in the first point above, it is necessary to use the metric to relate this physical length to a coordinate length, and the result of doing this gives what is called the $\bar\mu$ scheme.  If the coordinate and physical lengths are not related properly, it is well known in cosmological space-times that the resulting theory is not physically viable and does not have a good classical limit \cite{Ashtekar:2011ni}.

To avoid the difficulty of evaluating holonomies in the inhomogeneous radial direction (as described in the second point above), it is possible to consider the Schwarzschild interior only, whose geometry can be expressed in terms of the Kantowski-Sachs cosmological metric
\be
ds^2= - \left( \frac{R_{S}}{T} - 1 \right)^{-1} dT^2 + \left( \frac{R_{S}}{T} - 1 \right) dR^2 + t^2 d\Omega^2,
\ee
where $R_S = 2GM$ is the Schwarzschild radius, $d\Omega^2 = d\theta^2 + \sin^2\theta d\phi^2$, and $T$ is the radial coordinate that becomes time-like inside the horizon; this coordinate system is valid for $T \in [0, R_S)$.

For this choice of coordinates, the metric (and the connection) are independent of any spatial coordinate, and therefore holonomies in all spatial directions can be evaluated rather directly \cite{Modesto:2004xx, Ashtekar:2005qt}.  Despite this advantage, close to the horizon it is difficult to relate the coordinate length of a radial path in the $R$ direction with the physical length---as required by the $\bar\mu$ scheme---because $R$ becomes null at the horizon and $ds \to 0$.  Requiring that the physical length nonetheless be finite leads to unacceptably large quantum gravity effects near the horizon \cite{Bohmer:2007wi}.

Due to this problem, several alternate forms of holonomy corrections have been proposed \cite{Corichi:2015xia, Olmedo:2017lvt, Ashtekar:2018cay, Bodendorfer:2019cyv}, but what this discussion suggests is that the problem lies in the choice of coordinates that become null and that, to avoid these issues, it is necessary to use coordinates where there is a clean separation between time-like and space-like coordinates everywhere in the space-time.

As an aside, note also that the isometry between the Schwarzschild interior and the Kantowski-Sachs space-time depends on the dynamics of the space-time and is not guaranteed to hold once quantum gravity effects are included.  Among other possibilities, if there is an inner horizon in the LQG-corrected space-time then the Kantowski-Sachs metric could not describe the innermost region of the black hole lying within the inner horizon.  For an example of a modified gravity theory where Kantowski-Sachs is not isometric to the Schwarzschild interior see \cite{deCesare:2020swb}.

Therefore, in the following we will only consider choices of coordinates where the spatial coordinates always remain space-like, whether inside the horizon or out; the Painlev\'e-Gullstrand coordinates are one such example that we will use here.  The results in this paper can be adapted to several different coordinate choices, but the coordinates should always have the property that the spatial coordinates remain space-like everywhere.  Then, we will follow the general prescription for the $\bar\mu$ scheme that was first laid out in \cite{Bohmer:2007wi}, except here we will use coordinates that are everywhere space-like.

This approach has also been considered in some previous works \cite{Chiou:2012pg, Gambini:2020nsf}.  In this case, when considering coordinate choices such that the radial coordinate is always space-like, the challenge is to either evaluate holonomies in the radial direction (a difficult problem, in general) or to find a viable way to avoid doing so.  In the first work \cite{Chiou:2012pg}, the holonomies in the radial direction were evaluated in a `point-wise' fashion, this is an approximation where the path-ordering is dropped.  However, the constraints of the resulting theory did not close, showing that this approximation is not viable.  More recently, this problem was reconsidered using an Abelianized version of the constraints (see \cite{Gambini:2013ooa}) in which case only holonomies in the angular directions are needed to construct the Hamiltonian constraint operator \cite{Gambini:2020nsf}; this resulting theory is well-defined.

Here we will consider a complementary approach, where we will fix a gauge in the classical theory before introducing holonomy corrections, instead of using the Abelianized set of constraints as in \cite{Gambini:2013ooa, Gambini:2020nsf}.  Although the procedure that is followed here is slightly different, the end result is the same.  Specifically, we will gauge-fix the diffeomorphism constraint by imposing the areal gauge (which imposes that the prefactor to $d\Omega^2$ in the metric be $x^2$, with $x$ the radial coordinate).  This is a very simple gauge choice, which can always be chosen in spherical symmetry no matter the gravitational dynamics.  The areal gauge also has the additional property that the gauge-fixed Hamiltonian constraint only depends on the angular components of the Ashtekar-Barbero connection: as a result, the only holonomies that need to be evaluated are holonomies along paths that lie on spheres of constant radius---this avoids the difficulty of needing to evaluate holonomies in the radial direction and makes it possible to include holonomy corrections in a rather straightforward manner.  In Sec.~\ref{sec:CD}, we will go through this gauge-fixing procedure in detail, before continuing to the effective theory with LQG holonomy corrections in Sec.~\ref{sec:ED}.

\newpage

\section{Classical Theory}
\label{sec:CD}

The metric of any spherically symmetric space-time can be expressed in the form
\be \label{metric}
ds^2 = -N^2 dt^2 + f^2 (dx + N^x dt)^2 + g^2 d\Omega^2,
\ee
where the lapse $N(x,t)$, shift vector $N^x(x,t)$ and the functions $f(x,t), g(x,t)$ all depend on time $t$ and the radial coordinate $x$, while $d\Omega^2 = d\theta^2 + \sin^2 \theta \, d\phi^2$.  Note that we denote the radial coordinate by $x$, since it is not necessarily equal to the area radial coordinate $r$ that satisfies $A_r = 4 \pi r^2$, with $A_r$ being the surface area of the sphere at radius $r$.

\subsection{Basic Variables}
\label{sec:BV}

The spatial metric $q_{ab}$ can be rewritten in terms of the co-triads
\be
e_x^1 = f(x,t), \qquad e_\theta^2 = g(x,t), \qquad e_\phi^3 = g(x,t) \, \sin\theta,
\ee
with $q_{ab} = e_a^i e_b^j \delta_{ij}$.  The densitized triads are then given by $E^a_i = \sqrt{q} \, e^a_i$, with the triads $e^a_i$ satisfying $e_a^i e^b_i = \delta_a^b$ and $e_a^i e^a_j = \delta^i_j$, so
\be
E^x_1 = g^2 \sin\theta = E^a \sin\theta, \qquad E^\theta_2 = f g \, \sin\theta = E^b \sin\theta, \qquad
E^\phi_3 = f g = E^b,
\ee
with $E^a(x,t)$ and $E^b(x,t)$ capturing the degrees of freedom of the densitized triads.  The metric can now be rewritten as
\be \label{metric1}
ds^2 = -N^2 dt^2 + \frac{(E^b)^2}{{E^{a}}} \big(dx + N^x dt\big)^2 + E^a d\Omega^2.
\ee

The Ashtekar-Barbero connnection $A_a^i=\Gamma_a^i+\gamma K_a^i$ is the conjugate variable to the densitized triad, with the spin-connection given by
\be
\Gamma^i_a = \frac{1}{2} \, \epsilon^{i}{}_{jk} \, e^{bk} \left( \pd_{a} e^j_b - \pd_b e^j_a
+ e^{cj} e_{am} \pd_c e^m_b \right),
\ee
while the extrinsic curvature is $K_a^i = K_{ab} e^{bi}$, with $K_{ab} = \tfrac{1}{2} \mathcal{L}_t q_{ab}$, and $\gamma$ is the Barbero-Immirzi parameter.  Since the spatial metric is diagonal, so is $K_{ab}$ and we parametrize it by $a(x,t)$ and $b(x,t)$,
\be
\gamma K_x^1 = a, \qquad \gamma K_\theta^2 = b, \qquad \gamma K_\phi^3 = b \sin\theta,
\ee
while a short calculation gives
\be
\Gamma_\theta^3 = \frac{\pd_x E^a}{2E^b}, \qquad
\Gamma_\phi^1 = \cos\theta, \qquad
\Gamma_\phi^2 = -\frac{\pd_x E^a}{2E^b} \sin\theta,
\ee
all other components of the spin-connection are 0.

\subsection{Constraints and Dynamics}

The dynamics follow from the gravitational action \cite{Ashtekar:2004eh, Rovelli:2004tv, Thiemann:2007pyv},
\be
S = \int dt \int_\Sigma \left[ \f{\dot A_a^i E^a_i}{8 \pi G \gamma}
- N \mathcal{H} - N^a \mathcal{H}_a \right],
\ee
where dots denote derivatives with respect to $t$, the scalar constraint is
\be
\mathcal{H} = - \, \f{E^a_i E^b_j}{16 \pi G \gamma^2 \sqrt q} ~ \epsilon^{ij}{}_k \Big( F_{ab}{}^k - (1 + \gamma^2) \Omega_{ab}{}^k \Big),
\ee
and the diffeomorphism constraint is
\be
\mathcal{H}_a = \f{1}{8 \pi G \gamma} \, E^b_k F_{ab}{}^k.
\ee
Here the field strength is $F_{ab}{}^k = 2 \partial_{[a}^{} A_{b]}^k + \epsilon_{ij}{}^k A_a^i A_b^j$, while the spatial curvature is given by $\Omega_{ab}{}^k = 2 \partial_{[a}^{} \Gamma_{b]}^k + \epsilon_{ij}{}^k \Gamma_a^i \Gamma_b^j$.

Using the coordinate choices for spherical symmetry described in Sec.~\ref{sec:BV} and integrating over $d\Omega$ gives the symmetry-reduced action
\be
\label{eq:action}
S = \int dt \int dx \,  \left[ \f{\dot{a}\ea + 2\dot{b}\eb}{2G\gamma}
- N \mathcal{H} - N^x \mathcal{H}_x \right],
\ee
with the scalar constraint
\be
\mathcal{H} = - \, \frac{1}{2G\gamma} \Biggr[
\f{2ab\sqrt{E^a}}{\gamma} + \frac{E^b}{\gamma \sqrt{E^a}}(b^2+\gamma^2)
-\frac{\gamma (\partial_x{E^{a}})^2}{4 E^b\sqrt{{E^{a}}}}
-\gamma\sqrt{{E^{a}}}\partial_x\left(\frac{\partial_x{E^{a}}}{E^b}\right)\Biggr],
\ee
and the diffeomorphism constraint
\be
\label{eq:diffeo}
\mathcal{H}_x = \frac{1}{2G\gamma} \left(2 E^b \partial_x b - a \partial_x E^a \right).
\ee
Note that only the radial component of the diffeomorphism constraint is non-trivial once coordinates that are explicitly spherically symmetric have been chosen, as is the case here.

The action also shows that the symplectic structure of the symmetry-reduced theory is given by
\begin{align}
\{a(x_1),\ea(x_2)\}&=2G\gamma\,\delta(x_1-x_2), \\
\{b(x_1),\eb(x_2)\}&=G\gamma\, \delta(x_1-x_2).
\end{align}

Denoting $\mathcal{C}[N] = \int dx \, N \mathcal{H}$ and $\mathcal{D}[N^x] = \int dx \, N^x \mathcal{H}_x$, it is a straightforward, although long, calculation to verify that the constraint algebra (for the symmetry-reduced theory) is
\begin{align}
\{\mathcal{C}[N_1],\mathcal{C}[N_2]\} &= \mathcal{D} \left[ \frac{\ea}{(\eb)^2}
\left( N_1 \pd_x N_2 - N_2 \pd_x N_1 \right) \right], \\
\{\mathcal{D}[N^x_1],\mathcal{D}[N^x_2]\} &= \mathcal{D} [\left(N^x_2\pd_xN^x_1-N^x_1\pd_xN^x_2\right)], \\
\{\mathcal{C}[N],\mathcal{D}[N^x]\} &= -\mathcal{C}[N^x\pd_xN].
\end{align}

The equations of motion, determined by $\dot f = \{f, \mathcal{C}[N] + \mathcal{D}[N^x]\}$, are given by:
\be \label{eq:Eadot}
\dot E^a = \frac{2 Nb}{\gamma} \sqrt{E^a} + N^x \partial_x E^a,
\ee
\be \label{eq:Ebdot}
\dot E^b = \frac{N}{\gamma \sqrt{E^a}} (a E^a + b E^b) + \partial_x (N^x E^b),
\ee
\begin{align}
\label{eq:mota}
\dot{a} = & \, \frac{N}{2\gamma \sqrt{E^a}} \left[\frac{\eb}{E^a} (b^2+\gamma^2) - 2ab \right]
+ \frac{N \gamma}{2 \sqrt{E^a}} \left[\pd_x \left(\frac{\partial_x E^a}{\eb} \right)
- \frac{(\partial_x{E^{a}})^2}{4\ea\eb} \right] \nn \\
& + \gamma \, \pd_x \left(\frac{\partial_x(N\sqrt{{E^{a}}})}{\eb}\right)
-\f{\gamma}{2} \pd_x \left(N\frac{\partial_x{E^{a}}}{\eb\sqrt{{E^{a}}}}\right)
+ \partial_x (N^x a),
\end{align}
\be
\label{eq:motb}
\dot{b} = - \, \frac{N}{2\gamma \sqrt{E^a}} \left(b^2+\gamma^2\right)
- \frac{\gamma}{2} \left[ \frac{N}{\sqrt{E^a}} \left( \frac{\partial_x E^a}{2 E^b} \right)^2
- \partial_x \left(N \sqrt{E^a} \right) \frac{\partial_x E^a}{(E^b)^2} \right]
+ N^x \partial_x b.
\ee
After choosing a lapse and a shift, solutions to these equations of motion and the constraints $\mathcal{H} = 0$ and $\mathcal{H}_x = 0$ will give a metric \eqref{metric1} that satisfies the vacuum Einstein equations.

One example that will be relevant here is the Schwarzschild space-time for a black hole of mass $M$, expressed in Painlev\'e-Gullstrand coordinates for which
\be
N_{PG} = 1, \qquad N^x_{PG} = \sqrt{\frac{R_S}{x}},
\ee
where $R_S = 2 GM$ is the usual Schwarzschild radius.  Unsurprisingly, the solution is
\be
\label{eq:PGcoord}
\begin{split}
a_{PG} = \gamma\sqrt{\frac{R_S}{4x^3}}, \qquad E^a_{PG} = x^2, \\
b_{PG} = -\gamma\sqrt{\frac{R_S}{x}}, \qquad \eb_{PG} = x,
\end{split}
\ee
corresponding exactly to the Painlev\'e-Gullstrand metric.

\subsection{The Areal Gauge}

In order to simplify the passage to the effective Hamiltonian, and to avoid evaluating non-trivial path-ordered exponentials to calculate holonomies in the radial direction, we will perform a partial gauge-fixing known as the areal gauge.  This corresponds to setting $\ea=x^2$ (or, in the original metric \eqref{metric}, $g(x,t) = x^2$).  Importantly, this choice can be imposed without any reference to the equations of motion (indeed, this is typically done in textbook treatments of the Schwarzschild solution before even deriving the Einstein equations), so long as the surface area of spheres of constant $x$ increases monotonically with $x$, which can easily be checked once the solution is known.

The gauge-fixing condition $\chi = \ea - x^2 = 0$ is clearly second-class with the diffeomorphism constraint $\mathcal{H}_x$, and so can be used to gauge fix $\mathcal{H}_x$, giving
\be
\label{eq:GF}
\ea=x^2,\qquad a=\frac{\eb}{x} \, \pd_x b.
\ee
Then, requiring that this gauge be preserved by the equations of motion, i.e., $\dot \chi = 0$, imposes the condition that $\dot \ea = 0$ and therefore
\be
\label{eq:shlap}
N^x=-\frac{N b}{\gamma}.
\ee
Note that this implies that, after imposing the areal gauge-fixing condition, $b$ now appears in the metric through the shift vector $N^x$ which is no longer a Lagrange multiplier that can be freely chosen, but is fully determined once the lapse $N$ has been chosen.  On the other hand, the lapse remains a Lagrange multiplier that can be freely chosen and imposes the scalar constraint $\mathcal{H} = 0$.

This gauge significantly simplifies the action, which becomes
\be
\label{eq:SGF}
S_{GF} = \int dt \int dx \left[ \f{\dot{b}\eb}{G \gamma} - N \mathcal{H} \right],
\ee
with
\be
\label{eq:Careal}
\mathcal{H} =\frac{1}{2G\gamma} \biggr[\frac{3\gamma x}{E^b} - \frac{2\gamma x^2}{(E^b)^2}\pd_x E^b
- \frac{E^b}{\gamma x} \pd_x \left[x(b^2+\gamma^2)\right] \biggr].
\ee
Note that the symplectic term $\dot{a}\ea$ in \eqref{eq:action} becomes a total time derivative $(a \ea)^{\,}\dot{} \,$ since $\ea$ is independent of time, and so this term can be dropped.

The remaining Poisson bracket is
\be \label{pb}
\{b(x_1), \eb(x_2)\} = G \gamma \, \delta (x_1 - x_2),
\ee
and the constraint algebra also simplifies, becoming
\begin{align}
\{\mathcal{C}[N_1], \mathcal{C}[N_2]\} &= C\left[-\frac{1}{\gamma}\left(N_1 \pd_x N_2 - N_2 \pd_x N_1 \right) b \right] \nn \\
&= C\Big[ N^x_1 \pd_x N_2 - N^x_2 \pd_x N_1 \Big], \label{gr-cons}
\end{align}
using $\eqref{eq:shlap}$ to obtain the second relation.  The second form of the constraint algebra in the areal gauge will give some insight into what the correct form for the shift vector $N^x$ should be in the effective theory once LQG effects are included.

Finally, the equations of motion are
\be
\dot E^b =\frac{b}{\gamma x}\Big( N E^b - x \pd_x (NE^b) \Big),
\ee
\be
\dot b = \frac{\gamma N x}{2(E^b)^2} + \frac{\gamma x^2}{(E^b)^2} \pd_x N
-\frac{N}{2x\gamma} \pd_x (x b^2 + \gamma^2x ).
\ee
These equations can be obtained either by imposing the conditions \eqref{eq:GF} on the original equations of motion \eqref{eq:Eadot}--\eqref{eq:motb}, or by deriving them directly from the simplified scalar constraint \eqref{eq:Careal} via $\dot f = \{f, \int \! dx \, N \mathcal{H} \}$.  As expected, the solution for $N=1$ is exactly the Painlev\'e-Gullstrand metric.

\section{LQG Effective Dynamics}
\label{sec:ED}

The procedure to obtain the LQG effective dynamics for vacuum spherically symmetric space-times is to take the classical theory, described in Sec.~\ref{sec:CD}, then (i) replace the components of the Ashtekar-Barbero connection by holonomies, and (ii) include correction functions multiplying inverse powers of the densitized triad.  The first step is necessary since the basic operators in LQG are holonomies and areas (there is no operator corresponding to the connection itself), and it gives rise to `holonomy corrections'.  The second step arises because 0 is a discrete eigenvalue of the area operator in LQG, so there is no well-defined operator corresponding to, e.g., $1/\eb$; introducing well-defined operators corresponding to inverse powers of $E^a_i$ gives `inverse triad corrections'.

Here we will focus on holonomy corrections for two reasons.  First, in LQC the dominant quantum gravity effects comes from holonomy corrections: these are the source of the non-singular bounce, and it seems reasonable to expect that holonomy corrections will be dominant compared to inverse triad corrections in spherical symmetry as well.  Second, there is considerable ambiguity in the choice of inverse triad corrections, and in fact some choices of inverse triad operators in LQC do not generate any inverse triad corrections in the effective theory \cite{Singh:2013ava}.  Therefore, in the following we will assume that the inverse triad operator in the underlying quantum theory has an action such that there are no inverse triad corrections in the effective theory, and only consider holonomy corrections.

In LQC, the effective dynamics are known to provide an excellent approximation to the full quantum dynamics for states that are sharply peaked, and for which the expectation value for the spatial volume is always much larger than $\lp^3$ \cite{Taveras:2008ke, Rovelli:2013zaa}.  While it is not yet clear whether the effective dynamics will also provide a good approximation to the full quantum dynamics for black hole space-times, it seems likely that the arguments in \cite{Rovelli:2013zaa} can be generalized.  If this turns out to be the case, then the effective dynamics could be used to approximate the quantum dynamics of semi-classical states, at least for observables whose relevant physical length scale is much larger than $\lp$.  Based on this expectation, we will focus on the effective theory here, but we note that it is possible to construct the quantum theory following an analogous procedure to the one given in \cite{Gambini:2020nsf}.

\subsection{Effective Hamiltonian}
\label{sec:eff-ham}

To include holonomy corrections, it is necessary to replace in $\mathcal{H}$ the connection by holonomies.  This is, in the simplest cases, done by expressing the field strength $F_{ab}{}^k$ in terms of the holonomy of the Ashtekar-Barbero connection around a loop of minimal area $\Delta$, where the area gap $\Delta \sim \lp^2$ is the smallest non-zero eigenvalue of the area operator in LQG \cite{Ashtekar:2006wn}.

However, this procedure is not always viable in LQC when the spatial curvature is non-vanishing (as is the general case in spherical symmetry).  This is because the holonomy of $A_a^i$, evaluated around a loop of physical area $\Delta$, cannot be expressed as an operator on the LQC Hilbert space (to be specific, the holonomy cannot be written in terms of almost-periodic functions of the connection).  For the case of spherical symmetry, using the Cayley-Hamilton theorem it is possible to check that the holonomy of $A_a^i$ around a loop of minimal area is not almost periodic in $b$, and therefore a different approach is necessary.

This is a difficulty that has already been addressed in cosmological space-times with non-vanishing spatial curvature, and in this case what is known as the `K' loop quantization is preferred \cite{Vandersloot:2006ws, Singh:2013ava}.  For the case of spherical symmetry (and after imposing the areal gauge), this means replacing $b$ by holonomies of the extrinsic curvature 1-form $\gamma K_a^i$, evaluated in the $d\theta$ direction%
\footnote{We could equally well choose any path that follows a great circle, we choose $\phi=$ constant for simplicity.},
\be \label{hol}
h_\theta(\delta_b) = \exp \left( \int_0^{\delta_b} \gamma K_\theta^i \tau_i \, d\theta \right) 
= \cos \left( \f{\delta_b b}{2} \right) \mathbb{I} + 2 \sin \left( \f{\delta_b b}{2} \right) \tau_2.
\ee
where the $\tau^i$ are a basis in the (fundamental representation of the) $\mathfrak{su}(2)$ Lie algebra satisfying $\tau^i \tau^j = \tfrac{1}{2} \epsilon^{ij}{}_k \tau^k - \tfrac{1}{4} \delta^{ij} \mathbb{I}$, and $\mathbb{I}$ is the $2\times2$ identity matrix.

Then, to extract a scalar quantity from the $SU(2)$-valued expression \eqref{hol}, we replace
\be \label{sub-b}
b \to -2 \, \f{{\rm Tr} (h_\theta(2\delta_b) \cdot \tau_2)}{2\delta_b}.
\ee
Here the factor of 2 in $2 \delta_b$ is to ensure consistency between the `K' loop quantization and the loop quantization based on expressing the field strength in terms of holonomies \cite{Ashtekar:2009um}.

The remaining task is to determine the appropriate value for $\delta_b$.  The key heuristic argument from LQG, which guides the choice of $\delta_b$, is that the physical length of this edge should be given by $\sqrt\Delta$.  Since the holonomy was integrated along the edge with respect to the coordinate $\theta$, $\delta_b$ gives the coordinate length of the path, not the physical length.  The coordinate and physical lengths are simply related by the metric; for a path with constant $x$ and $\phi$ (and constant $t$, of course) the relation is just $ds = x \, d\theta$.  So, for the physical length to be $\sqrt\Delta$, the coordinate length must be taken to be
\be \label{delta}
\delta_b = \f{\sqrt\Delta}{x}.
\ee
(In general, if the areal gauge is not imposed then $ds = \sqrt{E^a} \, d\theta$ and $\delta_b = \sqrt{\Delta/\ea}$.)  This result is in agreement with what has earlier been argued in \cite{Bohmer:2007wi, Chiou:2012pg, Gambini:2020nsf} (up to an overall factor of $4\pi$ in some cases, which essentially implies a slightly different choice for $\Delta$).  Then, \eqref{sub-b} becomes
\be \label{brep}
b \to \f{x}{\sqrt\Delta} \sin\left( \f{\sqrt\Delta}{x} b \right).
\ee

It is now possible to construct the effective Hamiltonian by replacing all instances of $b$ in \eqref{eq:Careal} using \eqref{brep}, with the result
\be
\label{eq:modC}
\mathcal{H}^{(LQG)} = -\, \frac{1}{2G\gamma} \left[
\frac{E^b}{\gamma x}\pd_x\left(\frac{x^3}{\Delta}\sin^2 \frac{\sqrt{\Delta} \, b}{x} + \gamma^2 x \right)
-\frac{3\gamma x}{E^b} + \frac{2\gamma x^2}{(E^b)^2} \, \pd_x E^b \right].
\ee
A direct calculation of the Poisson bracket of the effective scalar constraint with itself gives the following constraint algebra,
\be \label{lqg-cons}
\{\mathcal{C}^{(LQG)}[N_1],\mathcal{C}^{(LQG)}[N_2]\} = C^{(LQG)}\left[- \, \frac{x}{\gamma\sqrt{\Delta}} \sin\frac{\sqrt{\Delta} \, b}{x} \cos \frac{\sqrt{\Delta} \, b}{x}
\left(N_1 \pd_x N_2 - N_2 \pd_x N_1 \right)\right].
\ee
Note that although the constraint algebra has changed compared to the classical form \eqref{gr-cons}, the constraint algebra for the effective scalar constraint is closed: there are no anomalies.

Next, it is necessary to update the areal gauge relation between the lapse and the shift, which is classically given by \eqref{eq:shlap}, by replacing $b$ by an appropriate expression in terms of holonomies.  A simple way to do this is in fact suggested by comparing the classical constraint algebra \eqref{gr-cons} and the constraint algebra in the effective theory \eqref{lqg-cons}: the choice
\be
\label{eq:modsh}
N^x = - \, \frac{N x}{\gamma \sqrt{\Delta}} \sin\frac{\sqrt{\Delta} \, b}{x} \cos\frac{\sqrt{\Delta} \, b}{x}
\ee
ensures that the constraint algebra for the effective theory will have exactly the classical form
\be
\{\mathcal{C}^{(LQG)}[N_1],\mathcal{C}^{(LQG)}[N_2]\} = C^{(LQG)}\left[ N^x_1 \pd_x N_2 - N^x_2 \pd_x N_1 \right].
\ee
This choice for the shift vector, although based on different arguments, is the same as in \cite{Gambini:2020nsf}.

As an aside, we mention that if a different modification for $b$ is preferred for the effective Hamiltonian, say $b \rightarrow f(x,b)$, then the constraint algebra will be $\{\mathcal{C}_f[N_1], \mathcal{C}_f[N_2]\} = C_f[-\gamma^{-1} (N_1 \pd_x N_2 - N_2 \pd_x N_1 ) (f \pd_b f) ]$, and by redefining the lapse-shift relation to be $N^x = - N (f \pd_b f) /\gamma$, the constraint algebra becomes identical with the classical case (And for a more general analysis of modified constraint algebras in spherical symmetry for the case of the diffeomorphism constraint not being gauge-fixed, see \cite{Aruga:2019dwq}).

For the choice \eqref{eq:modsh} for the shift vector, the effective metric will be
\be \label{metric2}
ds^2 = -N^2 dt^2 + \frac{(E^b)^2}{x^2} \big(dx + N^x dt\big)^2 + x^2 d\Omega^2.
\ee

Finally, from the scalar constraint and the basic Poisson bracket relation \eqref{pb}, the equations of motion for $E^b$ and $b$ are derived in the usual manner, giving
\be \label{eb-genN}
\dot E^b = - \, \frac{x^2}{\gamma \sqrt{\Delta}} \, \pd_x \left(\frac{N E^b}{x}\right)
\sin \frac{\sqrt{\Delta} \, b}{x} \cos \frac{\sqrt{\Delta} \, b}{x},
\ee
\be \label{b-genN}
\dot{b} =
\frac{\gamma N x}{2 (E^b)^2} \left( 1 + 2x \f{\pd_x N}{N} \right) - \frac{\gamma N}{2x}
-\frac{N}{2\gamma \Delta x} 
\pd_x \left( x^3 \sin^2 \frac{\sqrt{\Delta} \, b}{x} \right).
\ee

\subsection{Solution in Painlev\'e-Gullstrand Coordinates}

A stationary solution to the equations of motion and to the scalar constraint $\mathcal{H}^{(LQG)} = 0$ can easily be found in terms of Painlev\'e-Gullstrand-like coordinates for $N=1$.  For $N=1$, then $\dot \eb = 0$ implies that%
\footnote{If $\dot \eb = 0$, then either $E_b = x$, or $\sin \f{\sqrt\Delta \, b}{x} = 0$, or $\cos \f{\sqrt\Delta \, b}{x} = 0$.  In the second case, $\dot b = 0$ implies $E_b = x$ in agreement with the first case, while in the third case $\sin \f{\sqrt\Delta \, b}{x} = \pm1$ and $\dot b = 0$ gives $E_b^2 = \gamma^2 \Delta x^2 / (3 x^2 + \gamma^2 \Delta)$, which does not satisfy the scalar constraint. Therefore, only the first two cases are viable and both imply $\eb = x$.}
\be
\eb = x,
\ee
while $\dot b = 0$ gives
\be
b = \frac{x}{\sqrt{\Delta}} \arcsin \frac{C}{x^{3/2}},
\ee
where $C$ is a constant of integration.

It is immediately clear that $C=0$ gives Minkowski space, $ds^2 = -dt^2 + dx^2 + x^2 d\Omega^2$.  Note that there are no quantum gravity effects in this case, which is not surprising since the curvature is zero.

The black hole solutions are obtained for $C = - \sqrt{\gamma^2 \Delta R_S}$, where $R_S = 2GM$ is the Schwarzschild radius and $M$ is the mass of the black hole.  This is easily verified by considering the solution at large $x$, in which case $\arcsin (C/x^{3/2}) \approx C/x^{3/2}$ and the usual Painlev\'e-Gullstrand solution \eqref{eq:PGcoord} is recovered.

An important point, as already pointed out in \cite{Gambini:2020nsf}, is that this solution is only well-defined for
\be
x \ge x_{\rm min} = (\gamma^2 \Delta R_S)^{1/3}.
\ee
This lower bound on $x$ in vacuum space-times is not surprising given the following argument.  First, in spherically symmetric space-times there are no local gravitational degrees of freedom (gravitational waves), so a matter source is needed to generate any space-time curvature.  Second, studies in LQC show that quantum gravity effects due to holonomy corrections generate an upper bound on the possible energy density of any matter field.  Therefore, to generate a gravitational field corresponding to mass $M$, a matter field with density $\rho \sim M/R^3$ is needed, and if $\rho \le \rho_{\rm max} \sim \rho_{\rm Pl}$, then the matter field must extend to at least a radius of $\sim (M / \rho_{\rm Pl})^{1/3} \sim x_{\rm min}$.  This argument can be made precise, and shown to be exact, in the case that the matter field is pressureless dust field \cite{dust}.  So to describe the solution for $x < x_{\rm min}$, it is necessary to include matter fields.

For the vacuum part of the space-time, the shift vector is
\be \label{shift}
N^x = \sqrt{ \frac{R_S}{x} \left( 1 - \frac{\gamma^2 \Delta R_S}{x^3} \right)},
\ee
which gives the effective metric
\be
\label{eq:efle}
ds^2 = - \left(1 - \f{R_{S}}{x} + \f{\gamma^2 \Delta R_{S}^2}{x^4} \right) dt^2
+ 2 \, \sqrt{\frac{R_{S}}{x} \left(1 - \f{\gamma^2 \Delta R_S}{x^3} \right)} \, dt\,dx
+ dx^2 + x^2 d\Omega^2.
\ee
Note that in the limit $\Delta \rightarrow 0$, the effective metric tends to the classical Schwarzschild metric in Painlev\'e-Gullstrand coordinates, as expected.  Also, the condition $x \ge x_{\rm min}$ ensures that the shift $\eqref{shift}$ is well-defined for all $x \ge x_{\rm min}$.  Another interesting point is that in the limit $x \to x_{\rm min}$, the effective line element tends to the Minkowski metric.  As we shall see, this is because the repulsive quantum gravity effects exactly balance out the attractive classical gravitational force at $x = x_{\rm min}$.

\subsection{Curvature Scalars and Killing Horizons}
\label{s.curv-kh}

To understand the geometry underlying the effective metric \eqref{eq:efle}, it is useful to examine curvature scalars and look for horizons, with Killing horizons being particularly easy to find in stationary space-times.

In the following, to simplify the notation we will express the metric as $ds^2 = -F dt^2 + 2 N^x dt dx + dx^2 + x^2 d\Omega^2$, with
\be \label{FF}
F(x) \equiv 1 - \frac{R_S}{x} + \frac{\gamma^2 \Delta R_S^2}{x^4},
\ee
and $N^x$ given by \eqref{shift}, note that $F + (N^x)^2 = N^2 = 1$.

It is straightforward to calculate some simple curvature scalars for \eqref{eq:efle}, with the results
\be
R = - \, \frac{6 \gamma^2 \Delta R_S^2}{x^6}, \qquad
R_{\mu\nu}R^{\mu\nu} = \frac{90 \gamma^4 \Delta^2 R_S^4}{x^{12}},
\ee
\be
R_{\mu\nu\rho\sigma}R^{\mu\nu\rho\sigma} = \frac{12 R_S^2}{x^6} \left( 1 - \frac{10 \gamma^2 \Delta R_S}{x^3}
+ \frac{39 \gamma^4 \Delta^2 R_S^2}{x^6} \right).
\ee
Note that these expressions for the curvature scalars are exact.  Also, setting $\Delta=0$ in these equations gives the expected classical expressions, in particular $R_{\mu\nu} = 0$.  Further, as the lower bound $x_{\rm min}$ is approached, all of these curvature scalars approach a critical value that is independent of their mass, and which provides an upper bound to the amplitude of each curvature scalar in the vacuum region,
\be
\lim_{x \to x_{\rm min}} R = - \, \frac{6}{\gamma^2 \Delta}, \qquad
\lim_{x \to x_{\rm min}} R_{\mu\nu} R^{\mu\nu} = \frac{90}{\gamma^4 \Delta^2}, \qquad
\lim_{x \to x_{\rm min}} R_{\mu\nu\rho\sigma} R^{\mu\nu\rho\sigma} = \frac{360}{\gamma^4 \Delta^2}.
\ee
These upper bounds agree with the results obtained in \cite{Gambini:2020nsf} (up to overall factors of $4\pi$ due to what amounts to a different choice by \cite{Gambini:2020nsf} for $\Delta$ in \eqref{delta}).

\begin{figure}
\begin{center}
\includegraphics[width=0.65\textwidth]{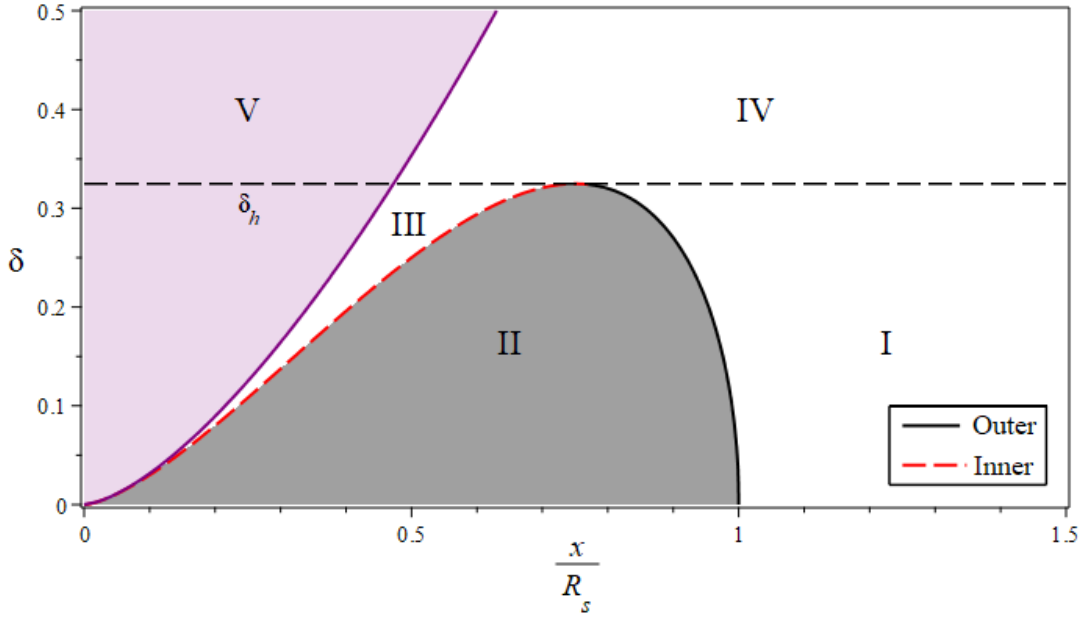}
\caption{This figure shows the location of $x_{\rm min}$ (solid purple line), $x_{\rm inner}$ (dashed red line) and $x_{\rm outer}$ (solid black line) as a function of $\delta = \gamma \sqrt{\Delta} / R_S$; the scale on the x-axis is in units of $x/R_S$.  
Region I corresponds to $x > x_{\rm outer}$, in Region II $x_{\rm inner} < x < x_{\rm outer}$, and in Region III $x_{\rm min} < x < x_{\rm inner}$.  For a black hole mass of $M <  M_\star$ there are no horizons, this corresponds to Region IV with $\delta > \delta_h$.  Finally, the effective solution is only valid for $x > x_{\rm min}$; in Region V the vacuum solution is not well-defined and it is necessary to include matter fields.  Note that in the limit of a large mass ($\delta \to 0$), $x_{\rm inner} \rightarrow x_{\rm min}$ while $x_{\rm outer} \to R_S$.} \label{horizons}
\end{center}
\end{figure} 

Next, in an explicitly stationary space-time like this one, $\xi^\mu = (1,0,0,0)$ is necessarily a Killing vector field and the Killing horizons are located where $\xi^\mu \xi_\mu = 0$, which corresponds to $F = 0$.  What is interesting here is that (for $M \gg m_{\rm Pl}$) there are two Killing horizons%
\footnote{$F=0$ gives a fourth-order polynomial in $x$; two roots are always complex and for $M \gg m_{\rm Pl}$ the other two roots are real and distinct.  As will be explored next, there is a limiting case $M = M_\star$ where there is one repeated real root, and for $M < M_\star$ all four roots are complex, in which case there is no Killing horizon.}:
an outer Killing horizon near $x = R_S$, and an inner Killing horizon just outside $x_{\rm min}$.

Specifically, to leading order in $\Delta / R_S^2$ the outer Killing horizon is located at
\be \label{outer}
x_{\rm outer} = R_S - \frac{\gamma^2 \Delta}{R_S} + O \left( \f{\Delta^{4/3}}{R_S^{5/3}} \right),
\ee
while the inner horizon is located at
\be
x_{\rm inner} = (\gamma^2 \Delta R_S)^{1/3} + \frac{1}{3} \left( \frac{\gamma^4 \Delta^2}{R_S} \right)^{1/3}
+ O \left( \f{\Delta^{4/3}}{R_S^{5/3}} \right),
\ee
note that the first term is exactly $x_{\rm min}$.  The location of the Killing horizons as a function of the black hole mass is shown in Fig.~\ref{horizons}.

\begin{figure}
\centering
\begin{tikzpicture}
  \fill[gray!10!white] (0.02,0.05) rectangle (2,5.95);
  \draw[->, very thick](0,0)--(0,6.2);
  \draw[->, very thick](0,0)--(11.5,0);
  \draw(2,0)--(2,6);
  \draw(3,0)--(3,6);
  \draw(8,0)--(8,6);
  \draw[fill] (10,3) circle [radius=0.05];
  \draw[dashed, very thick](10,3)--(9.3,3.7);
  \draw[dashed, very thick](10,3)--(10.7,3.7);
  \draw[fill] (8,3) circle [radius=0.05];
  \draw[dashed, very thick](8,3)--(8,4);
  \draw[dashed, very thick](8,3)--(7,3);
  \draw[fill] (5.5,3) circle [radius=0.05];
  \draw[dashed, very thick](5.5,3)--(4.8,3.7);
  \draw[dashed, very thick](5.5,3)--(4.8,2.3);
  \draw[fill] (3,3) circle [radius=0.05];
  \draw[dashed, very thick](3,3)--(3,4);
  \draw[dashed, very thick](3,3)--(2,3);
  \node[left] at (0,6){\large{\textit{t}}$\,$};
  \node[below] at (11.5,-0.1){\large{\textit{x}}};
  \node[below] at (0,0){$0$};
  \node[below] at (2,-0.1){$x_{\rm min}$};
  \node[below] at (3,-0.1){$x_{\rm inner}$};
  \node[below] at (8,-0.1){$x_{\rm outer}$};
\end{tikzpicture}
\caption{Schematic diagram for $M \gg m_{\rm Pl}$ showing the behaviour of the lightcone in the presence of multiple horizons. The shaded area is outside the domain of our solution and the innermost region between $x_{\rm min}$ and $x_{\rm inner}$ is a shell of thickness $\sim (\gamma^4 \Delta^2 / R_S)^{1/3}$.}
\label{cones}
\end{figure}
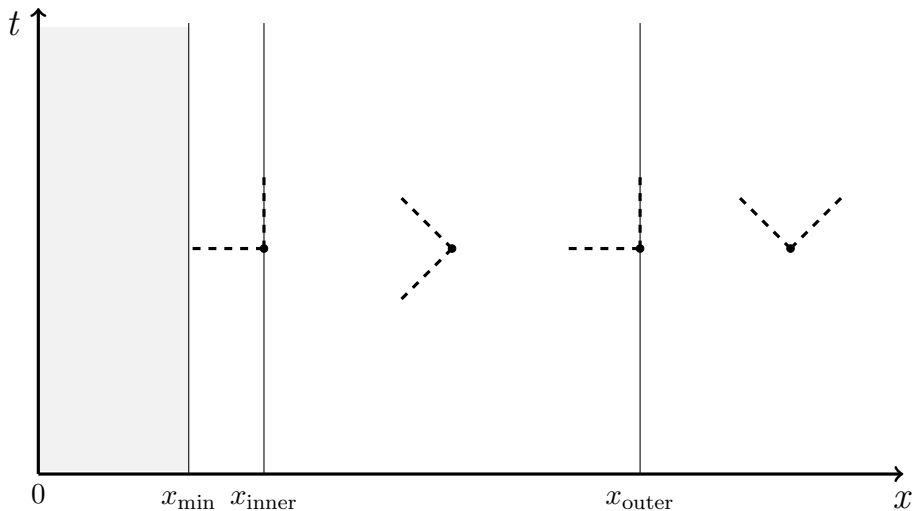

This shows that the outer Killing horizon is located (up to small quantum corrections) at the classical horizon $R_S$, while the interior horizon is a new feature of the quantum geometry that lies within the region where the space-time curvature is Planckian.  As depicted in Fig.~\ref{cones}, this shows that there is a thin region inside the black hole where the lightcone flips again and outgoing null rays begin to expand (with respect to the coordinate $x$) once again.  Note that the presence of an interior horizon is analogous to what occurs in Reissner-Nordstr\"om black holes, although the new term in the Reissner-Nordstr\"om metric with charge $Q$ goes as $G Q^2 / x^2$ while here the quantum gravity correction in the effective metric is proportional to $\Delta R_S^2 / x^4$.

This result also emphasizes the importance of studying the full space-time rather than using the classical isometry between the Schwarzschild interior and the Kantowski-Sachs space-time, which implicitly assumes that there is no interior horizon.  Also, note the existence of an interior horizon is a necessary condition for a transition to occur from a black hole collapse to an expanding white hole solution \cite{BenAchour:2020gon}; for details in how such a transition is realized in this effective framework for the case that the matter field is pressureless dust, see \cite{dust}.

The above results for the locations of the two horizons assumes $M \gg m_{\rm Pl}$, but if $M$ is sufficiently small there may be only 1 or 0 Killing horizons.  The limiting case occurs for $M_{\star} = 8 \gamma \sqrt{\Delta} / \sqrt{27} G$, when there is exactly one Killing horizon, while if $M < M_\star$ then there are no Killing horizons at all.  Although it is likely that the effective description fails for small $x < R_S$ in a space-time with such a small mass, the absence of Killing horizons in this case is nonetheless interesting as it suggests that a minimal mass is required to form a black hole, with a (Killing) horizon---if $M < M_\star$, the space-time is indeed curved by the mass but not sufficiently for a horizon to form.  This is very different from the situation in classical general relativity, where there is always a horizon surrounding a sufficiently compact matter source.  Note that this also suggests that elementary particles with $m < m_{\rm Pl}$ (like electrons, say) cannot form a black hole alone; rather, many elementary particles must be packed in a sufficiently small region for a black hole to form.

\subsection{Geodesics and Apparent Horizons}

Further insight into the effective geometry of the LQG-corrected black hole can be obtained by studying geodesics.  For the sake of simplicity we will consider radial motion only, but it is straightforward to extend these results to include rotational motion as well.

\begin{figure}
\begin{center}
\includegraphics[width=0.72\textwidth]{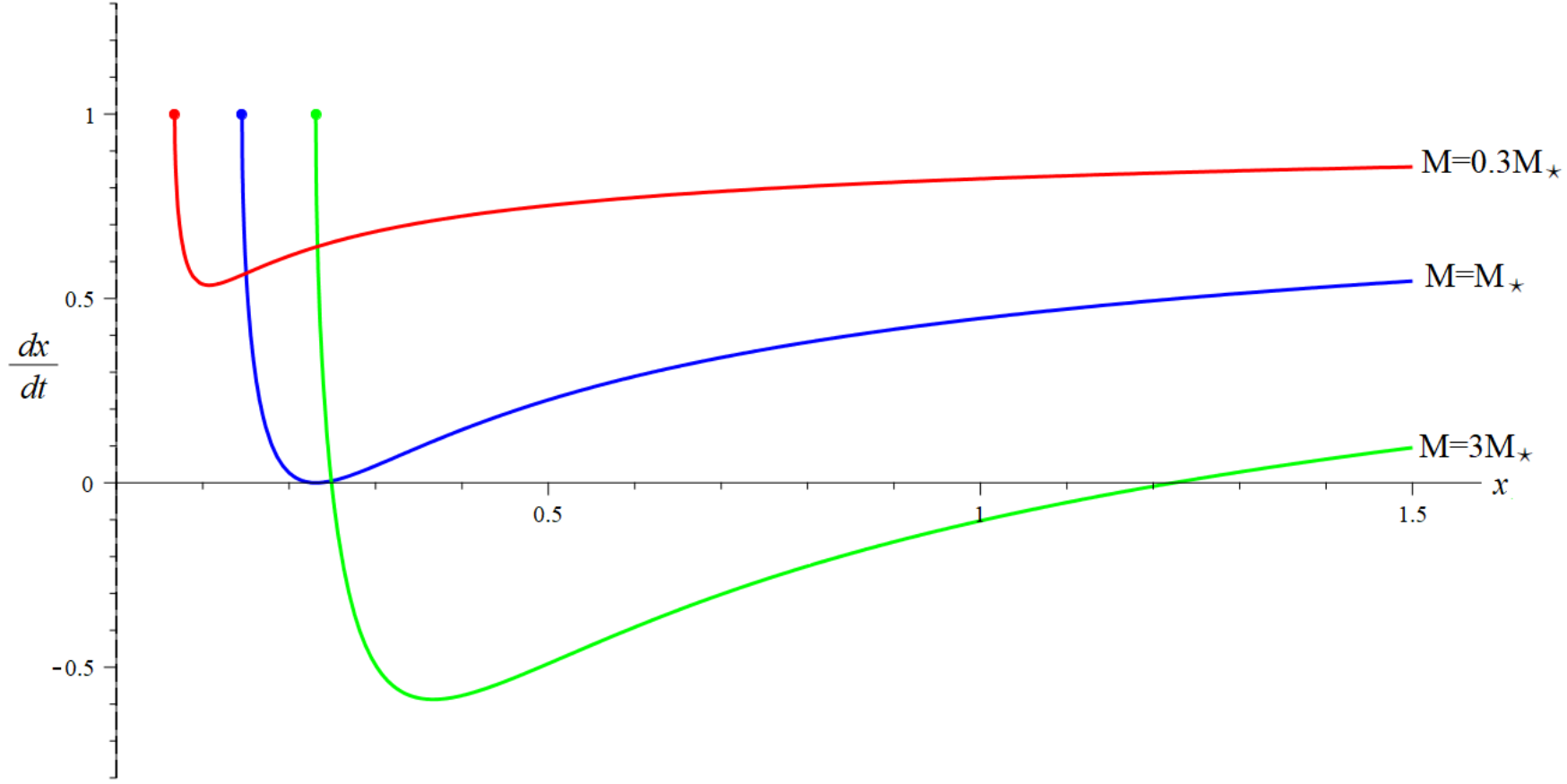}
\end{center}
\caption{Behaviour of outgoing radial null geodesics, $dx/dt = 1-N^x$, for the effective metric with shift $\eqref{shift}$ for black holes of different mass.  These three curves are characteristic of the different possible behaviours that the outgoing null rays may exhibit depending on the mass of the black hole compared to $M_{\star}$.  Here the red curve corresponds to $M=0.3 M_{\star}$, the blue curve to $M=M_{\star}$, and the green to $M=3 M_{\star}$.  The abrupt end to each curve corresponds to $x=x_{\text{min}}$ for each of the three black hole masses; the vertical dashed black lines depicts the radial coordinate value at which this happens for each mass.  In these plots we set $\gamma=G=1$ and $\Delta=10^{-2}$.} \label{null}
\end{figure}

Radial geodesics satisfy
\be \label{geo-gen}
-\epsilon=-F(x)\dot{t}^2+2N^x\dot{t}\dot{x}+\dot{x}^{2} \, ;
\ee
for time-like geodesics, $\epsilon=1$ and the dots denote derivatives with respect to proper time $\tau$, while for null geodesics $\epsilon=0$ and dots denote derivatives with respect to an affine parameter $\lambda$.

For time-like geodesics, it is convenient to use the conserved energy associated with the time-like Killing vector $\xi^\mu = (1,0,0,0)$ to isolate $\dot x$; specifically,
\be \label{cons}
E \equiv \xi_{\mu} \dot{x}^{\mu} = -F(x) \dot{t} + N^x \dot{x}.
\ee
Combining this with $\epsilon = 1$, the geodesic equation \eqref{geo-gen} simplifies to
\be \label{geo}
- F(x) = -(N^x \dot{x} - E)^2 + 2 N^{x} (N^x \dot{x} - E) \dot x + F \dot{x}^2,
\ee
giving
\be 
\frac{dx}{d\tau}=\pm\sqrt{ E^2 - 1 + \frac{R_{S}}{x} \left( 1 -\frac{\gamma^2\Delta R_S}{x^3} \right)} \, .
\ee
Note that in the case $E=1$, corresponding to a particle that starts at rest at infinity, this particle will again have $\dot x = 0$ at $x = x_{\rm min}$.  This is another way to see that $x = x_{\rm min}$ is the location where the quantum gravity repulsive effects cancel out the classical gravitational attraction.

For null geodesics the calculation is even simpler.  Since $\epsilon = 0$, dividing \eqref{geo-gen} by $\dot t^2$ gives
\be
0 = -F(x) + 2 N^x \frac{dx}{dt} + \left( \frac{dx}{dt} \right)^2,
\ee
which has the solution
\be  \label{nullgeo}
\frac{dx}{dt} = -N^x \pm 1.
\ee
For $dx/dt = -N^x - 1$, the ingoing null rays always have decreasing $x$, but the situation is a little more complicated for the outgoing rays with $dx/dt = 1 - N^x$ which will depend on the location of the zeros of $1 - N^x$; unsurprisingly these correspond exactly to the Killing horizons found in Sec.~\ref{s.curv-kh}.

For $M \gg m_{\rm Pl}$, the $x$ position of the outgoing rays will increase for $x > x_{\rm outer}$ and $x < x_{\rm inner}$, but decrease for $x_{\rm inner} < x < x_{\rm outer}$.  On the other hand, if $M < M_\star$, then the outgoing null rays will satisfy $dx/dt > 0$ everywhere.  This is depicted in Fig.~\ref{null}; once again the behaviour is analogous with that of a Reissner-Nordstr\"om black hole.

\begin{figure}
\begin{center}
\includegraphics[width=1\textwidth]{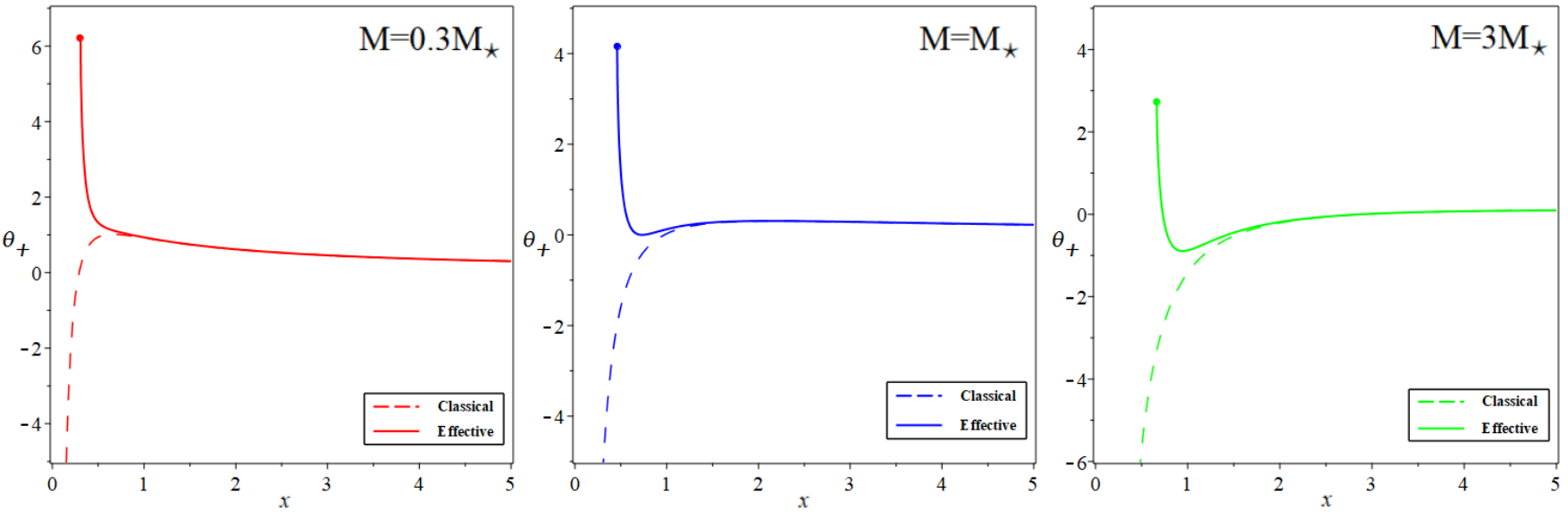}
\end{center}
\caption{Comparison of the outgoing null expansion $\theta_{+} = 2 x^{-1} (1-N^x)$ for the effective metric $\eqref{eq:efle}$ (solid line) compared to the classical limit $\Delta \rightarrow 0$ (dashed line).  The three cases, from left to right, correspond to: (i) a space-time with no apparent horizon (left), (ii) a space-time with one sphere $S$ that is marginally trapped sphere (where $\theta_+ = 0$) but no region with $\theta_+ < 0$ (middle), and (iii) a space-time with two apparent horizons (right).  In the first two cases, $\theta_{+} \geq 0$ throughout the entire space-time while in the large mass case of $M > M_\star$, $\theta_+$ is negative for $x_{\rm inner} < x < x_{\rm outer}$.  Each of these examples are qualitatively different from the classical case where, in all examples, at $x = R_S$ the expansion $\theta_+$ becomes negative and diverges to $-\infty$ as $x\rightarrow 0$.  Note that $\theta_+$ for the effective metric stops at $x = x_{\rm min}$, whose location is denoted by the dotted vertical line. In the plots we set $\gamma=G=1$ and $\Delta=10^{-2}$.} \label{exp}
\end{figure} 

Next, it is possible to determine whether there are any apparent horizons by considering congruences of null geodesics.  Due to spherical symmetry, it is sufficient to consider congruences that are orthogonal to the surface of concentric 2-spheres $S$ defined by constant $x$ and $t$.  Denoting the tangent vector to the outgoing null geodesics by $\ell^\mu = (1, 1-N^x, 0, 0)$, then the other linearly independent null vector that is also orthogonal to $S$ is $k^\mu = (1, -1-N^x, 0, 0)$, which is the tangent vector to ingoing null geodesics.  Here the overall normalization of these two vectors fields is such that $\ell^\mu k_\mu = -2$, so the hypersurface metric for $S$ is given by
\be
h_{\mu\nu} = g_{\mu\nu} + \f{1}{2} \Big( \ell_\mu k_\nu + k_\mu \ell_\nu \Big).
\ee
The outgoing and ingoing expansions are respectively
\be
\theta_+ = h^{\mu\nu} \nabla_\mu \ell_\nu, \qquad \theta_- = h^{\mu\nu} \nabla_\mu k_\nu,
\ee
and a short calculation gives
\be
\theta_+ = \frac{2}{x} ( 1-N^x ), \qquad \theta_- = - \frac{2}{x} (1 + N^x ).
\ee
The standard definition of a trapped surface $S$ is one where both expansions $\theta_\pm$ are negative, and the boundary of the total trapped region is called the apparent horizon---in this case, since $\theta_- < 0$ for all $x$, the apparent horizon corresponds to the surfaces where $\theta_+ = 0$.  Interestingly, for $M \gg m_{\rm Pl}$, in addition to the usual outer boundary to the trapped region, there is also an interior boundary and there are therefore two apparent horizons.  As expected, these apparent horizons are located at precisely the same location as the Killing horizons, $x_{\rm inner}$ and $x_{\rm outer}$.  The expansion $\theta_+$ is plotted in Fig.~\ref{exp} for different $M$ and compared to the classical result.

\begin{figure}
\begin{center}
\includegraphics[width=0.6\textwidth]{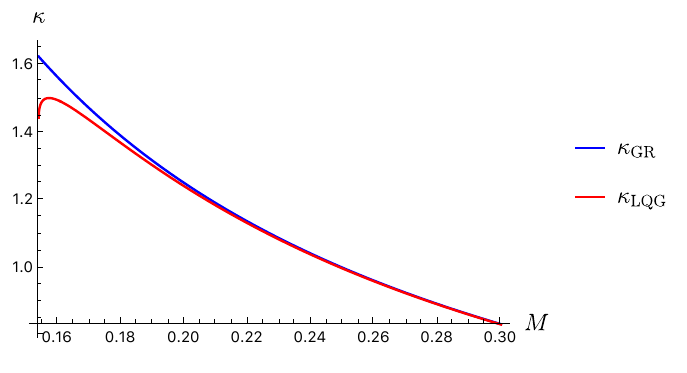}
\end{center}
\caption{This plot shows the (outer) horizon surface gravity $\kappa$ as a function of the black hole mass $M$ for small $M$;  the smallest mass shown here is $M = M_\star$, as there are no horizons for $M < M_\star$.   Note that the slope of the curve is positive for small $M$ close to $M_\star$, but $\kappa$ rapidly tends to the classical result $\kappa = 1 / 2R_S$.
In this plot we set $\gamma=G=1$ and $\Delta=10^{-2}$.} \label{surface-gravity}
\end{figure}

It is straightforward to calculate the surface gravity at the outer horizon,
\be
\kappa = \f{R_S}{2 x_{\rm outer}^2} - \f{2 \gamma^2 \Delta R_S^2}{x_{\rm outer}^5}.
\ee
In the case that $M \gg m_{\rm Pl}$, then the outer horizon is given by \eqref{outer} and the surface gravity, to leading order in $\Delta$, is given by
\be
\kappa = \f{1}{2 R_S} - \f{\gamma^2 \Delta}{R_S^3} + O \left( \f{\Delta^{4/3}}{R_S^{11/3}} \right).
\ee
It is also interesting to examing the surface gravity for smaller masses which is shown in Fig.~\ref{surface-gravity}; of course, since there is no horizon for $M < M_\star$, a surface gravity can be associated to a horizon only for $M \ge M_\star$.  It is interesting to note that the slope of $\kappa(M)$ is positive for small $M$, so the specific heat of black holes becomes positive for sufficiently small mass (assuming the black hole thermodynamics correspondence between surface gravity and temperature continues to hold in this setting).

Finally, for large $M$ and keeping only the leading order LQG correction, the black hole thermodynamics relation for these effective (non-rotating, zero charge) black holes is slightly modified to
\be
\kappa \, \delta \! A_{\rm outer} = 8 \pi G \left(1 - \f{2 \gamma^2 \Delta}{R_S} \right) \delta \! M,
\ee
suggesting that, not too surprisingly, quantum gravity effects will generate some departures from semi-classical expectations based on quantum field theory on a classical background.  A more detailed exploration of this topic is left for future work.

\subsection{Other Coordinate Systems}

While we have so far used the Painlev\'e-Gullstrand coordinate system, it is also possible to express the stationary solution in terms of other coordinate systems.

Leaving $N$ free in \eqref{eb-genN} and \eqref{b-genN}, requiring $\dot \eb = 0$ gives
\be
N = \f{x}{\eb}
\ee
which can then be substituted into $\dot b = 0$, with the result
\be
\f{x^2}{\Delta} \sin^2 \f{\sqrt\Delta \, b}{x} = \gamma^2 \left( \f{x^2}{(\eb)^2} - 1 + \f{R_S}{x} \right),
\ee
where the constant of integration has been chosen to obtain the correct classical limit at large $x$.

To make contact with \cite{Gambini:2020nsf}, we will now consider the specific example
\be
N = \f{1}{\sqrt{1 + R_S/x}},
\ee
for which it follows that
\be
\eb = x \sqrt{1 + R_S / x}, \qquad
\f{x^2}{\Delta} \sin^2 \f{\sqrt\Delta \, b}{x} = \f{\gamma^2 R_S^2}{x^2 (1 + R_S/x)},
\ee
and so the resulting effective line element is
\begin{align} \label{efle-gop}
ds^2 = & \, - \left( 1 - \f{R_S}{x} + \f{\gamma^2 \Delta R_S^4}{x^6 (1 + R_S/x)^2} \right) dt^2
+ \left( 1 + \f{R_S}{x} \right) dx^2 \nn \\ & \quad
+ 2 \, \f{R_S}{x (1 + R_S/x)} \sqrt{1 - \f{\gamma^2 \Delta R_S^2}{x^4 (1 + R_S/x)}} ~ dx \, dt
+ x^2 d\Omega^2.
\end{align}
This is precisely the effective metric found in \cite{Gambini:2020nsf}, up to some small discretization effects which are not included in the analysis here.  This shows that the different approaches followed here and in \cite{Gambini:2020nsf} give the same effective metric, and are consistent with each other.

While the effective line elements \eqref{eq:efle} and \eqref{efle-gop} are both solutions of the effective equations of motion \eqref{eb-genN} and \eqref{b-genN}, these two effective metrics are not related by coordinate transformations, as would be the case in classical general relativity.  Rather, there is a quantum deformation to this classical symmetry; we leave a determination of the precise properties of this deformation for future work.  We emphasize that although the classical symmetry is deformed due a modification of the structure function in the effective algebra $\eqref{lqg-cons}$, the modified constraint algebra does close, thus ensuring the covariance of the effective model considered here.

One way to verify that the two metrics are not related by a coordinate transformation is to compare curvature scalars.  While $R$, $R_{\mu\nu} R^{\mu\nu}$ and $R_{\mu\nu\rho\sigma} R^{\mu\nu\rho\sigma}$ are all slightly different for the effective line elements \eqref{eq:efle} and \eqref{efle-gop}, nonetheless these curvature scalars have a nearly identical behaviour (especially for large $M$, with differences only becoming apparent near $x_{\rm min}$ for small $M$) and in fact have exactly the same upper bound that in both cases is reached at $x = x_{\rm min}$.

The fact that the space-time geometry depends on the coordinates---or, in other words, is observer-dependent---is (at least in hindsight) not surprising.  It is well known in the context of quantum field theory on curved space-times that different observers see different states: one may observe the quantum vacuum, while another (at the same location but with a relative acceleration) sees a thermal state.  Something similar appears to occur here: the quantum gravity corrections to the classical metric are observer-dependent (although some quantum gravity effects, like the presence of a Planck-scale bound on curvature scalars, appear to be observer-independent).  An in-depth study of this effect is left for future work.

\section{Summary and Discussion}
\label{sec:sum}

In this paper we constructed an effective framework to study quantum gravity holonomy effects in vacuum spherically symmetric space-times, and studied the stationary solutions to the effective theory.  By imposing the areal gauge, it was possible to implement the $\bar\mu$ loop quantization scheme; in an important sign of the robustness of these results, this gives results in perfect agreement with the $\bar\mu$ loop quantization based on the Abelianized version of the constraints \cite{Gambini:2020nsf}.

We explored the geometry of the solution mostly in terms of the effective line element expressed in Painlev\'e-Gullstrand-like coordinates, and found that quantum gravity effects: (i) slightly shift the location of the outer horizon from $x = R_S$ by a term of the order $\Delta/R_S$, (ii) showed that the vacuum solution only holds for $x \ge x_{\rm min} = (\gamma \Delta R_S)^{1/3}$, with the implication that the presence of matter is necessary at smaller $x$ to curve the space-time, and (iii) there is now an inner horizon located just outside $x_{\rm min}$ where the outgoing expansion of radial null geodesics becomes positive again.  (In this effective space-time, Killing horizons and apparent horizons are the same, so we simply refer to `horizons'.)  Note that the presence of an outer and inner horizon occurs in many models of non-singular black holes, including the well-known Bardeen model \cite{Bardeen:1968}.  Further, in agreement with \cite{Gambini:2020nsf}, the curvature scalars $R$, $R^{\mu\nu} R_{\mu\nu}$ and $R^{\mu\nu\rho\sigma} R_{\mu\nu\rho\sigma}$ are all bounded by quantum gravity effects, with each bound depending only on $\gamma^2 \Delta \sim \lp^2$ and independent of $M$.

Also, while this may lie outside the regime of validity of the effective description, it is nonetheless interesting to point out that the effective theory predicts that for sufficiently small $M \lesssim m_{\rm Pl}$ there will not be any horizon at all: although the mass will curve the space-time in the usual way far from the source, the gravitational field will never be strong enough to generate a trapped region, even for $x \le R_S$; this provides a quantum gravitational counterexample to the hoop conjecture for sufficiently low mass objects.  

The static space-time solution that we derived as a solution to the effective scalar constraint corresponds to an eternal (vacuum) black hole.  To describe physical black holes, it will be important to extend these results to include matter fields and to study the process of the formation of a black hole.  In addition, by including dynamical effects, it will be possible to study the problem of mass inflation, which appears to indicate that inner horizons in an eternal black hole are unstable \cite{Brown:2011tv, Alesci:2011wn, Carballo-Rubio:2018pmi} (although this conclusion has recently been challenged, see \cite{Bonanno:2020fgp}).  More generally, to properly understand the properties of astrophysically relevant black holes, it will be essential to include matter and allow for fully dynamical space-times.

Importantly, the effective framework developed here can be extended to include matter fields, and in particular it is quite straightforward to include a pressureless dust field \cite{dust}.  Then, it is possible to study black hole collapse; for example, it can be shown that in the Oppenheimer-Snyder collapse model the dynamics of the interior of the `star' are given by exactly the LQC effective Friedman equation, and therefore the star bounces at the LQC critical density $\rho_c \sim \rho_{\rm Pl}$ and then starts to expand, much like a white hole \cite{dust}.  This model is a first step towards a more complete analysis of quantum gravity effects in a dynamical black hole space-time, starting from the collapse of an in-falling matter field.  Interestingly, it provides an explicit realization, derived from an effective LQG description of the full black hole space-time, that shows how quantum gravity effects can generate a transition from a collapsing black hole to an expanding `white hole', as suggested in \cite{Rovelli:2014cta}.

There also remain several other important open questions, in addition to the inclusion of matter and the study of the dynamics of evolving black hole space-times starting from their initial formation.  In particular, it is important to explore the relation of the effective metric expressed in terms of different coordinates, and to understand precisely how quantum gravity effects will differ depending on the observer.  While these questions are by now quite well understood for quantum field theory on curved space-times, this is not the case for quantum gravity effects, even in relatively simple effective theories like the one considered here.  Finally, it will also be important to go beyond the effective description in order to study other quantum gravity effects in black holes, most notably Hawking radiation and the black hole information loss problem.

\begin{acknowledgments}

We thank Viqar Husain for helpful discussions.
This work was supported in part by the Natural Sciences and Engineering Research Council of Canada.

\end{acknowledgments}

\raggedright

\end{document}